# Review

# Use of machine learning in geriatric clinical care for chronic diseases: a systematic literature review

Avishek Choudhury , Emily Renjilian, and Onur Asan *

School of Systems and Enterprises, Stevens Institute of Technology, Hoboken, New Jersey, USA

*Corresponding Author: Onur Asan, PhD, School of Systems and Enterprises, Stevens Institute of Technology, 1 Castle Point Terrace, Hoboken, NJ 07030, USA: oasan@stevens.edu

## ABSTRACT

**Objectives:** Geriatric clinical care is a multidisciplinary assessment designed to evaluate older patients' (age 65 years and above) functional ability, physical health, and cognitive well-being. The majority of these patients suffer from multiple chronic conditions and require special attention. Recently, hospitals utilize various artificial intelligence (AI) systems to improve care for elderly patients. The purpose of this systematic literature review is to understand the current use of AI systems, particularly machine learning (ML), in geriatric clinical care for chronic diseases.

**Materials and Methods:** We restricted our search to eight databases, namely PubMed, WorldCat, MEDLINE, ProQuest, ScienceDirect, SpringerLink, Wiley, and ERIC, to analyze research articles published in English between January 2010 and June 2019. We focused on studies that used ML algorithms in the care of geriatrics patients with chronic conditions.

**Results:** We identified 35 eligible studies and classified in three groups: psychological disorder ($n = 22$), eye diseases ($n = 6$), and others ($n = 7$). This review identified the lack of standardized ML evaluation metrics and the need for data governance specific to health care applications.

**Conclusion:** More studies and ML standardization tailored to health care applications are required to confirm whether ML could aid in improving geriatric clinical care.

## INTRODUCTION

According to the US Census Bureau, by 2050, the geriatric population will increase to 88.5 million.[1,2] Typically, geriatric patients suffer from multiple ailments and chronic conditions.[3] Multimorbidity, *the presence of multiple chronic diseases in a patient*, affects majority of the geriatric population.[4,5] The complicated geriatric syndromes result in poor health outcomes,[6] disability, mortality, and institutionalization rates.[7] Figure 1 illustrates some of the major concerns related to geriatric patients and their care providers. One of the most significant challenges in caring for geriatric patients is developing an accurate and fast diagnosis.[8,9] Such patients bring complex health histories and clinical scenarios into health care practices that make it essential to emphasize *how to improve patient-care outcomes for this population.*[10–12]

Geriatric patients, due to limited cognitive (30% of elderly patients have dementia[13]) or physical ability, typically fail to present their illness and symptoms.[14] Their assessment differs from a typical medical evaluation and is more challenging due to their limited cognitive and physical ability. Geriatric patients are also prone to inadequate nutritional intake because of factors, including polypharmacy, decreased mobility, and physiological changes.[15] Often such patients experience deterioration of mental[16] or physical health during their hospital stay, even if they recover

**LAY SUMMARY**

Patients above the age of 65, also referred to as geriatric or older patients, often suffer from multiple chronic diseases and require special medical attention. Their limited functional, physical health, and cognitive ability make it challenging for geriatricians to diagnose and develop the necessary plan of care. Recently, hospitals have begun to utilize various artificial intelligence (AI) systems to improve care for older patients. AI comes with many clinical promises, and in this review, we explored studies which used machine learning (ML), a component of AI, to identify psychological disorders, especially Alzheimer's disease, among older patients. Studies reported positive outcomes and often used algorithms such as support vector machines and deep-learning algorithms. Despite all the positive impacts of ML models, as indicated in the literature, we identified some issues regarding the use of ML for diagnosing chronic diseases among older patients. The two major issues discussed in this study are the lack of standardized ML evaluation metrics and the need for data governance specific to health care applications. Resolving the identified issues in this review may improve ML usage and facilitate care for older patients.

from the primary chief of concern for the admission.[17] Studies have also shown that hospitalized geriatric patients have a 60-fold increased risk of developing permanent disabilities[18] making them more susceptible to other adverse ailments.[19–21] Studies have shown that at about 30% of geriatric patients with an acute medical concern exhibit a gradual decline in their ability to adhere to Activities of Daily Living (ADLs).[22–26] Since ADLs are prerequisites to self-care and independent living,[22,26,27] the inability of elderly patients to perform ADLs has resulted in hospitalization-associated disability[17] such as cognitive impairment and delirium.[28,29] The literature portrays the comprehensive geriatric assessment as a time-consuming process requiring a multidisciplinary approach.[30] Unfortunately, physicians have limited time in their visits to examine geriatric patients with multiple concerns.[31] Consequently, the increasing burden of clinical documentation, inefficient technology,[32–34] and shortage of physicians (geriatricians)[35] also influence care quality (incomplete diagnosis).[14]

In medicine, artificial intelligence (AI) comes with promises to offer better prevention, diagnosis, and treatment. AI technologies have cancer detection,[36–42] disease management using robotics,[43–52] and other patient safety factors.[53–63] AI technologies have helped clinicians make decisions[64–66] and improved drug development[67–69] and patient-care monitoring.[70–73] AI has recently outperformed human performance in some domains.[74] Screening tools for dementia are being developed for detecting early cognitive impairments[75–79] and other geriatric health problems such as fall risks[80] and urinary tract infections among patients with dementia.[16]

With the gradual growth of AI applications in the health care industry and the availability of data,[81,82] it is reasonable to assume a positive impact of AI on geriatric patients. However, several limitations have been reported concerning these AI-based tools. Dr. Cabitza and colleagues[84] rightly cautioned and acknowledged the potential risks of AI in health care that can occur due to *the uncertainty of health care data* and the non-*explainability of complex deep-learning algorithms*. Therefore, the implicit notion that existing AI technologies can improve geriatric health outcomes, in our view, is a questionable assumption. Therefore, it is crucial to understand current practices using AI to assist clinicians in geriatrics care. In this study, we conducted a systematic literature review to understand how AI has been used for the care of geriatric patients with chronic ailments.

## MATERIALS AND METHODS

This systematic review is reported according to the Preferred Reporting Items for Systematic Reviews and Meta-Analysis guidelines (PRISMA).[85]

## Scope

In this study, we refer to chronic diseases as *ailments that persist for an extended period*. We defined elderly patients as *individuals over the age of 65 and have one or more chronic illnesses*.

AI is broadly defined as a computer program (that operates with predefined rules and data-driven models) that is capable of making intelligent decisions.[86] AI tools are technologies, whether it be a computer application or health care device that can analyze data, present hidden information, identify risks in patient health, and communicate diagnoses.[87] Through the use of the MeSH database, the definition of "artificial intelligence" was better understood in the different components that apply to it. It includes applications such as machine learning (ML), computer heuristics, neural networks, robotics, expert systems, knowledge bases, fuzzy logic, and natural language processing. For the scope of AI within this study, this term has been limited to applications involving only ML algorithms. *ML enables computers to utilize labeled (supervised learning) or unlabeled data (unsupervised learning) to identify hidden information or make classification about the data without explicit programming*.[70,87] However, discussing the subfield of AI is beyond the scope of this review.

We narrowed the scope of AI to ML in particular due to its noteworthy societal impact on the health care domain.[88] Moreover, ML is a significant component AI system used in health care.[70] With the increasing amount of data within the healthcare industry, the prevalence of implementing ML is gaining momentum.[89] A large amount of data are available from electronic health records (EHRs), which contains both structured and unstructured data,[88] and ML methods can allow computers to learn from EHR data and develop predictions by identifying hidden patterns.[87,90]

## Information sources and search

We used eight databases: PubMed, WorldCat, MEDLINE, ProQuest, ScienceDirect, SpringerLink, Wiley, and ERIC, to search for peer-reviewed articles. The search criteria were limited to articles that were published in English within the last 10 years (between January 1, 2010 and June 2019).

We identified our search terms with the help of the librarian. We developed two sets of keywords to encompass the eligibility criteria. The first set of keywords used "*artificial intelligence*" OR "*machine learning*" OR "*deep learning*" injunction with an AND operator and the terms "*elderly patients*" OR "*older adults*." The second set of keywords consisted of MeSH terms using the PubMed MeSH database. The used terms were "*disease attributes*", "*aged*," and "*artificial intelligence*." The MeSH term "*disease attributes*" was used and then sorted to ensure all articles, including chronic illnesses along with the other keywords, were included. The term included topics such as "*acute disease*," "*asymptomatic diseases*," "*catastrophic illness*," "*chronic disease*," "*convalescence*," "*critical illness*," "*disease progression*," "*disease resistance*," "*disease susceptibility*," "*diseases in twins*," "*emergencies*," "*facies*," "*iatrogenic disease*," "*late-onset disorders*," "*neglected diseases*," "*rare diseases*," and "*recurrence*."

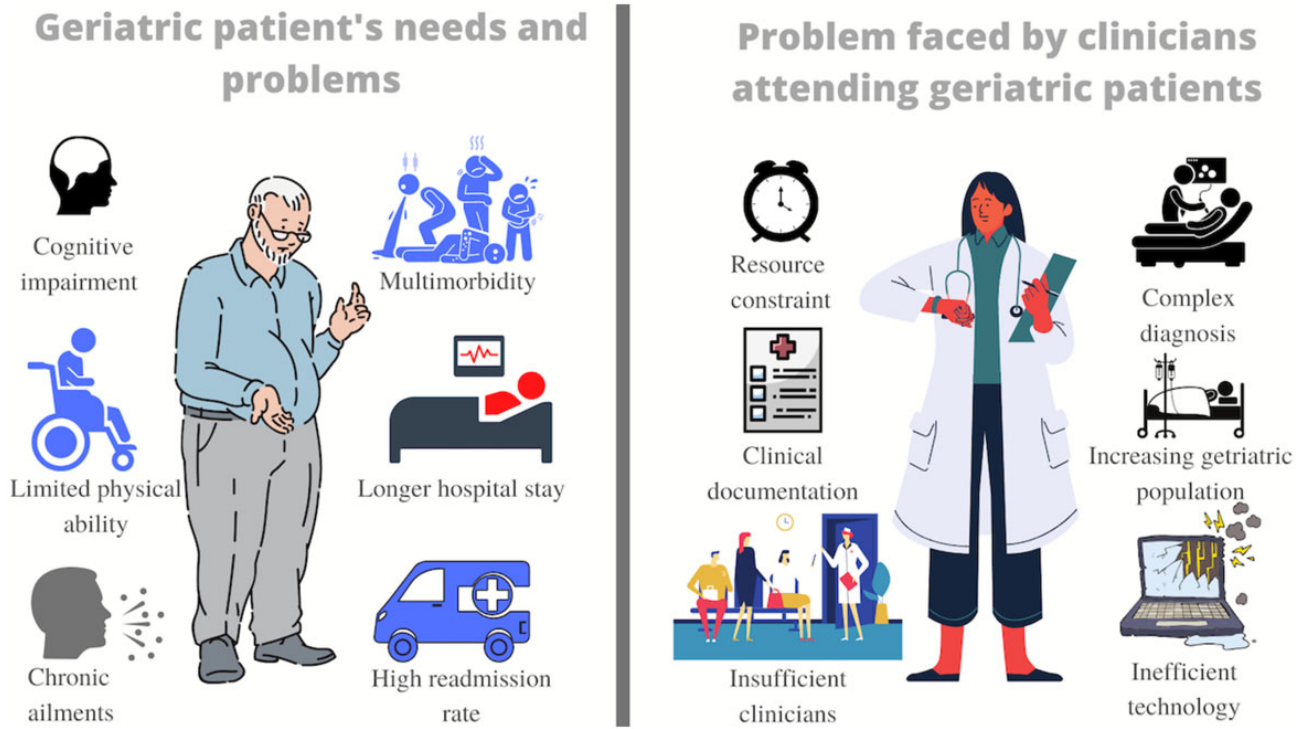

**Figure 1.** Graphical illustration of geriatric needs and clinician's problems.

MeSH terms are organized in a tree-like hierarchy, with more specific (narrower) terms arranged beneath broader terms. By default, PubMed includes in the search all narrower terms; this is called "exploding" the MeSH term. Moreover, the inclusion of MeSH terms optimizes the search strategy. All of the articles that fit our inclusion criterion were analyzed to make sure the disease they were targeting was chronic before continuing forward.

### Inclusion and exclusion criteria

This study focused on peer-reviewed publications satisfying the following three conditions:

- Implementation of ML techniques to address chronic ailment in elderly patients.
- Reporting or discussing changes in studied patient outcomes/conditions
- The study only involves geriatric patients.

We excluded any study that did not report a measurable patient outcome, opinion/review papers, qualitative perception papers, and studies involved patients other than the geriatric population.

### Study selection and quality assurance

All three authors together analyzed potential publications for their eligibility. We initially screened by reading abstracts and titles. Then, we read the full text to identify eligible articles for our inclusion criteria. Clarification on article inclusion was discussed between team members as necessary when a reviewer was unsure of whether to include or exclude a given article. We resolved all discrepancies by requiring consensus from all three reviewers and the librarian to minimize any selection bias.

### Data collection process

Data were collected using a data abstraction form to record standardized information from each selected article. We recorded the author, title, objective, method, health issue, gender, and findings of all publications. We then categorized the publications into different chronic illnesses, sources of data, and ML algorithms used by them.

## RESULTS

The set of queries illustrated earlier, returned *407* publications in PubMed, *104* publications in WorldCat, *85* in Medline, *57* in ProQuest, *21* in ScienceDirect, *13* in SpringerLink, *8* in Wiley and *1* in ERIC, so a total of *696* papers (Figure 2). We removed *262* duplicate publications (using EndNote X9.3.2). The authors screened the remaining *434* studies by reading abstracts and titles. Three hundred eighty-eight publications that did not meet our inclusion criteria were removed, and *46* papers were shortlisted for full screening. Eleven papers that did not meet our inclusion criteria were removed after the comprehensive screening of full text. The remaining *35* articles matched our inclusion criteria and were included in the systematic review with consensus from all three reviewers. The out-come of this process is *35* publications within the targeted scope and inclusion criteria.

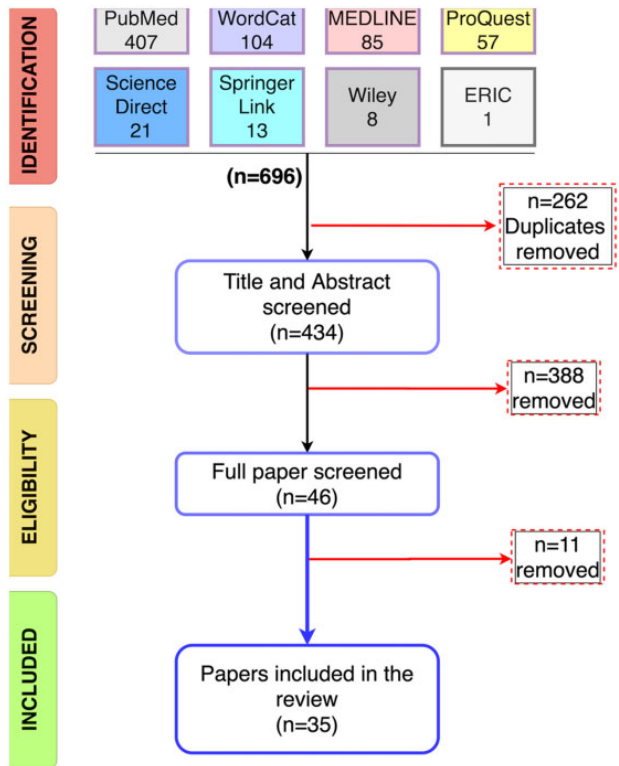

**Figure 2.** PRISMA selection procedure.

## Characteristic of the studies

The characteristics of the studies are reported in Supplementary Table S1 with an overview of the information including author, title, the objective of the paper, the ML model used in the article, the condition/ailment, participant characteristics specifically gender, and findings of the studies (see Supplementary Table S1). Another table summarizes all 30 ML algorithms, identified in the review, with the details of model performance measures and consecutively demonstrates the heterogeneity in ML reporting (see Supplementary Table S2).

Figure 3 shows the major sources of data and types of algorithms used by different studies in our review. Figure 4 presents a general introduction to the kinds of models identified in our review and the frequency of their usage by various studies. The general explanations of model types are based on the standard developed under the cognizance of the Consumer Technology Association (CTA) R13 Artificial Intelligence Committee,[91] The Royal Society of Britain,[92] and the author's knowledge. The support vector machine was the most frequently used model, followed by deep-learning methods and decision trees. Note the purpose of these figures (Figures 3 and 4) is not to provide an exhaustive technical insight but to highlight important issues relevant to ML models in the studied applications. We also categorized studies based on the type of health condition (Table 1) and reported the data types (Table 2), respectively.

## Findings in the text

"*Effective prevention and early detection*" is one of the major standards of care for geriatric patients developed by Luchi et al in 2003.[128] This standard mandate physician (geriatricians) to critically evaluate the screening recommendations and selectively apply them to achieve the patient's individual health care goals by following the "*Individualized Health Maintenance Protocol*".[128] In other words, geriatricians must tailor health maintenance measures to in-dividual patient's functional needs, health care goals, and preferences. Moreover, geriatric care may differ from standards set for younger adults and children. As geriatric needs may evolve with increasing age, geriatricians should periodically address and revise their care plan based on the varying medical conditions. However, no studies in our review developed or implemented ML models that can provide personalized care. Studies also did not account for changes in patient health conditions. The majority of the studies that employed ML simplified the diagnostic problems to binary classifications. For instance, studies analyzed blood samples to classify inflammatory bowel disease and coronary artery disease.[110] Another study used OCT images, retrieved from the HARBOR clinical trial, to classify chordal neovascularization and geographic atrophy.[122] Similarly, studies used MRI scans to classify MCI converters and MCI non-converters[102] and AD and healthy controls.[103] Such binary classification ignores the fact that chronic diseases can co-exist (multiple ailments) or exist in multiple layers of severity. For example, AD can be decomposed into further classes like "Light Autism," "Severe Autism," etc.[129] Therefore, the binary classification might not necessarily account for the multimorbidity and the complexity of geriatric health problems.

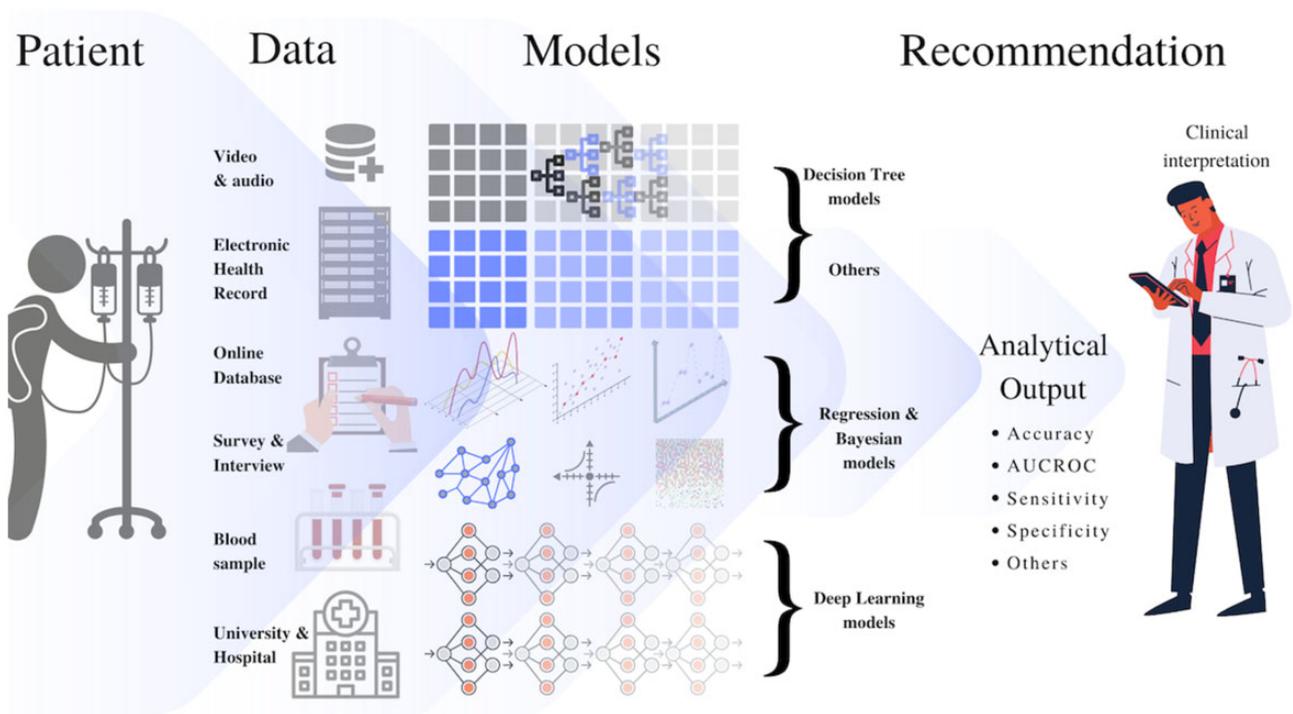

**Figure 3.** Type of data source and the types of models identified in the review.

| Models | General explanation |
|---|---|
| Deep Learning models (n = 13) 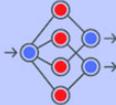 | These models are inspired by human biological nervous systems, although there are differences pertaining to the structural and functional properties of biological brain. The elementary constituents of these models are neurons, which can be considered as functions that receive inputs and produce an output that is a weighted sum of the inputs fed through an activation function. |
| Regression models (n=11) 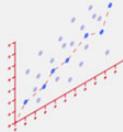 | It is a technique used in AI and ML, in which the application continuously estimates the relationship between outputs through the use of numeric values. |
| Decision Tree models (n=12) 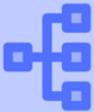 | These models uses a decision tree to navigate from observations about an item to conclusions about the item's target value. Models where the target variable can take discrete values are called classification trees; in these tree structures, leaves represent class labels and branches represent conjunctions of features that lead to those class labels. Random forests is a type of ensemble method consisting of a multitude of decision trees at training time and yielding the class that is the mode of the classes or mean prediction of the individual trees. |
| Bayesian models (n=7) 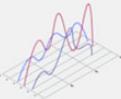 | Bayesian statistics is a branch of statistics where quantities of interest are treated as random variables, and one draws conclusions by analyzing the posterior distribution over these quantities given the observed data. In machine learning these models are broadly used to solve the problem of parameter estimation and model comparison. |
| Other models (n=22) Support vector machine 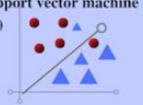 | These models sorts data into two categories using an optimal hyperplane. It is trained with a series of data already classified into two categories, building the model as it is initially trained. The task of an SVM algorithm is to determine which category a new data point belongs in. |
| others (n=6) 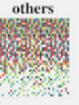 | This category consists of machine learning algorithms including k-nearest neighbor, King Lu algorithm, tariff, etc. These models, like all other machine learning algorithms, have the ability to learn and change without programming an explicit mathematical model for mapping input to output. |

**Figure 4.** General introduction to the type of models identified in the review and their frequency of use.

Most studies claim the derivation of an ML method for disease diagnosis. In contrast, in most cases, the researchers have merely adopted existing ML algorithms and implemented them separately on their dataset. To optimize ML measures such as sensitivity, specificity, AUROC, or classification accuracy, studies have commonly strived to differentiate between healthy and unhealthy participants or identifying certain chronic diseases and risks. Different versions of the input dataset (different features, different databases, and different data types) were trained to maximize the measures mentioned above, and studies recommended the ML model that yielded the best performance results. This means if a different dataset with variations in features is used, a new possible system will be recommended. Thus, the models derived earlier will not be valid or will not yield the same predictive results[129] (lack of reproducibility). Therefore, most classification and diagnostic systems' predictive powers in all the current studies rely heavily on the input features besides sampling and data quality. Most studies in our review dealt with chronic geriatric ailments (Table 2). The models were not evaluated against any procedural standard (clinical gold standard) or tested using real-time data over a period of time. Therefore, these studies can be seen as promising research, but not as a complete classification system or diagnostic method for geriatric ailments.

One major challenge we observed is the unavailability of benchmarked datasets for the use of ML. The majority of the studies trained their models using the datasets (Table 2) that varied in ways researchers process or collected them (initial dataset) according to the original investigation's requirements. Due to the absence of a standardized dataset, we have identified discrepancies among study results (difference in ML performance despite using data from the same source). Often data analysis requires extensive preprocessing to make the data suitable for ML algorithms, and different algorithms require specific types of preprocessing or data type.[129] Therefore, studies used variety of approaches (depending on the algorithms employed) to process different versions of the same dataset (different sizes or different types), which makes it challenging for others to reproduce the results or validate its reliability in a clinical setting. For instance, Chenet et al[99] preprocessed all MRI and PET data by performing anterior commissure-posterior commissure (AC-PC) correcting. The AC-PC corrected images were resampled to 256 × 256 × 256 voxels. The skull-stripping method was used to review MRI images manually, whereas Ortiz et al[119] did not implement AC-PC correction. Rather this study resized the MRI images to 121 × 145 × 121 voxels and PET images to 79 × 95 × 68 voxels.

**Table 1.** Disease classification

| Disease name | Disease type | Number of publications |
| --- | --- | --- |
| Mild cognitive impairment | Psychological disorder | 22 |
| Alzheimer's disease | | |
| Creutzfeldt Jacob disease | | |
| Autism spectrum disorder | | |
| Depression | | |
| Schizophrenia | | |
| Parkinson's disease | | |
| Age-related macular degeneration | Eye diseases | 6 |
| Diabetic retinopathy | | |
| Glaucoma | | |
| Geographic atrophy | | |
| Angina pectoris | Other ailments | 7 |
| Asthma | | |
| Chronic obstructive pulmonary disease | | |
| Cirrhosis | | |
| Hearing loss | | |
| Osteoarthritis | | |
| Rheumatoid arthritis | | |
| Inflammatory bowel disease | | |
| Hepatitis C virus infection | | |
| Coronary artery disease | | |

## DISCUSSION

With the invasion of ML in health care, automated, and data-driven clinical decision support systems[130,131] have gained popularity. To our knowledge, this is the first systematic review portraying the role and influence of ML on geriatric clinical care for chronic diseases. Our study identified the lack of ML standardization including (1) *heterogeneity in ML evaluation* (which evaluation metric should be reported?), and (2) *lack of a framework for data governance* (What kind of data are suitable for training a particular ML model? What should be the sample size of training data?). Development of some AI/ ML standards, such as (1) *ISO/IEC CS 23053 Framework for artificial intelligence (AI) systems using machine learning (ML)*[132] and (2) *ISO/WD TR 22100-5 Safety of machinery—relationship with ISO 12100-part5: implications of embedded artificial intelligence—machine learning,*[133] is under progress at ISO,[134] the leading standards body. But these ongoing standards efforts are primarily tailored to address ethical concerns. Although international standards to support ethical and policy goals are essential, there remains a risk that these standards may fail to address the concerns identified in our review.

### Heterogeneity in ML evaluation

We acknowledge that different algorithms might require different metrics for evaluation. Our review identified *heterogeneity in ML evaluation*. As reported in Supplementary Table S2, studies using the same (similar) algorithm used different evaluation metrics (or different combinations of measures) to determine ML performance. Many studies in our review only reported accuracy measures. However, AUROC is considered to be a superior metric to classification accuracy.[135,136] AUROC measures are beneficial for providing a visual representation of the relative trade-offs between the true positives and false positives of classification regarding data distributions. Albeit, in the case of unbalanced data sets, the ROC curves may provide an overly optimistic view of an algorithm's performance.[137] In such situations, the precision-recall curves can provide a more informative representation of performance assessment.[138] As sensitivity (recall), and precision give slightly different information, and they should be interpreted differently.[137] The abstract measures used to evaluate ML algorithms are not clinically meaningful, and understanding the ML evaluation metric requires technical knowledge.[139] In a research setting (where time and urgency are not drivers to action), interpreting ML metrics such as accuracy, sensitivity, and specificity can be perused theoretically. On the other hand, in a clinical setting, decisions made on an inappropriate metric(s) (decisions based on accuracy only) can lead to unintended consequences. Besides, making clinical decisions based on appropriate parameters might not necessarily ensure patient safety. ML algorithms trained on flawed (biased, incorrect subjective) data or data obtained from an insufficient sample can still generate misleading outcome measures.

Most of the studies in our review trained their models using historical data. Historical data retrieved from medical practices contains health care disparities in the provision of systematically worse care for vulnerable groups than for others.[140] In the United States, historical health care data reflect a payment system that rewards the use of potentially unnecessary care and services and may be missing data about uninsured patients.[139] Therefore, reliable data, along with standardized ML, may facilitate geriatric care.

**Table 2.** Data source and number of participants

| References | Data source | Data type | No. of patients |
| --- | --- | --- | --- |
| [93] | Sensing technologies | Signals | 97 |
| [94] | DIARETDB 1 | Fundus autofluorescence (FAF) images | – |
| [95] | Self[a] | Self-reported mood scores | 40 |
| [96] | Self[a] | Self-reported scales and Neurologist based scales | 410 |
| [97] | Self[a] | Video | 27 |
| [98] | Japanese Alzheimer's Disease Neuroimaging Initiative (J-ADNI) | MRI scans | 231 |
| [99] | • Alzheimer's Disease Neuroimaging Initiative Database (ANDI) <br> • Australian Imaging, Biomarker & Lifestyle database (AIBL) | MRI scans | 1,302 |
| [100] | Retinologist scanned the patient's eyes | Optical coherence tomography (OCT images) | 38 |
| [101] | Electroencephalographic (EEG) data | Spatial invariants of EEG data | 143 |
| [102] | Alzheimer's Disease Neuroimaging Initiative Database (ANDI) | MRI scans | 100 |
| [103] | Alzheimer's Disease Neuroimaging Initiative Database (ANDI) | MRI scans | 202 |
| [104] | National Social Life, Health, and Aging Project Wave 2 data (NSHAP) | Physical health and illness, medication use, cognitive function, emotional health, sensory function, health behaviors, social connectedness, sexuality, and relationship quality | 3377 |
| [105] | Diagnostic Innovations in Glaucoma (DIGS) study | Optical coherence tomography (OCT images) | 121 |
| [106] | • Accelerometers (sensors) <br> • Patient's medical record | Signals | 52 |
| [107] | Osteoarthritis Initiative database (OAI) | MRI scan | – |
| [108] | • Diagnostic Innovations in Glaucoma (DIGS) study <br> • African Descent and Glaucoma Evaluation Study (ADAGES) | Optical coherence tomography (OCT images) | 418 |
| [109] | Randomized controlled trials | Scales and questionnaires | 284 |
| [110] | Biobank—(UKSH tertiary referral center) | Blood samples (RNA) | 114 |
| [111] | Population Health Metrics Research Consortium (PHMRC) Study | Questionnaire | 1200 |
| [112] | Memory Clinic located at the Institute Claude Pompidou in the Nice University Hospital | Audio recording | 60 |
| [113] | Taiwanese mental hospital | Paper-based medical records | 185 |
| [114] | GenBank database | Nucleotide sequence | 17 |
| [115] | Alzheimer's Disease Neuroimaging Initiative Database (ANDI) | MRI scans | 1618 |
| [116] | Degenerative Diseases at Laboratrio de Biologia Molecular do Centro de Oncohematologia Pedítrica | Cognitive test results | 151 |
| [117] | The Magna Graecia University of Catanzaro and Regional Epilepsy Center, Reggio Calabria; Neurologic Institute "Carlo Besta," Milano; Neurologic Institute, University of Catania | Electroencephalographic (EEG) data | 195 |
| [118] | Self[a] | Blood samples (DNA extraction) | 648 |
| [119] | Alzheimer's Disease Neuroimaging Initiative Database (ANDI) | MRI scans | 275 |
| [120] | Alzheimer's Disease Neuroimaging Initiative Database (ANDI) | MRI scans | 72 |
| [121] | A longitudinal case-control study. Subjects were recruited via posted flyers from the local community | MRI scan | 178 |
| [122] | HARBOR clinical trial (ClinicalTrials.gov identifier: NCT00891735) | Optical coherence tomography (OCT images) | 1097 |
| [123] | Alzheimer's Disease Neuroimaging Initiative Database (ANDI) | MRI scans | 113 |
| [124] | Self-captured using Spectralis, Heidelberg Engineering, Heidelberg, Germany | Fundus autofluorescence (FAF) images | – |
| [125] | Two more extensive studies at Washington State University | Interview, testing, and collateral medical information | 582 |
| [126] | Chang Gung Memorial Hospital | Clinical Dementia Rating (CDR) and the Mini-Mental State Examination (MMSE) score | 52 |
| [127] | Alzheimer's Disease Neuroimaging Initiative Database (ANDI) | MRI scans | 281 |

[a]Indicates that the data were collected by the researcher or author of that paper (not from any database or prior study).

## Need for data governance
Our review shows the need for data governance which has also been acknowledged by the Royal Society of Great Britain.[92] The overarching goal of data governance includes not only the aspects of le-gal and ethical norms of conduct but also conventions and practices that govern the *collection*, storage, *use*, and transfer of data.[92] ML algorithms and their outcomes highly depend on data.[141] Their properties, such as reliability, interpretability, and responsibility, rely on the quality of data they have been trained on. Most of the studies we reviewed used data from databases (that store complete and standardized data for research purposes) and observational studies. It is challenging to determine whether the results of observational studies are unbiased and true. Models trained on such data are ideal for research purposes (model development). Still, they might not work as efficiently in a clinical environment (model validation and implementation) where data are unstructured, incomplete (missing values), and biased.[142]

EHRs are one of the primary sources of data in hospitals. Very few studies in our review used data from hospitals. EHR or paper-based data stored and collected by hospitals are prone to bias due to the under and over-representation of specific patient populations.[87,143] Besides, different institutions record patient information differently; as a result, if ML models trained at one institution are implemented to analyze data at another institution, this may result in errors.[140] Studies that used blood samples (DNA and RNA) to train their model are also prone to bias since the majority of sequenced DNA comes from people of European descent.[144,145] Therefore, in the context of ML and healthcare, data governance should address questions as to whether a specific data set (collected from a small sample; old; collected for research purposes; etc.) or type of data (subjective; patient-reported; digital data; etc.) can be used for a particular purpose (diagnosis; prognosis; clustering; mining; etc.).

To improve geriatric care, models must not only be developed but also integrated into clinical workflow. Our review did not identify any study that integrated their model into clinical workflow. Given a lack of interoperability standards mentioned above, for a model to work, it must be able to interface with the data within different EHRs. Unlike data obtained from online research databases or research institutions (as observed in our review), each EHR contains varying data structures (often not compatible with other systems), creating a significant challenge to model deployment.[146] Recently, the Department of Health and Human Services, led by the Office of the National Coordinator for Health Information Technology (ONC), released the draft *2020-2025 Federal Health IT Strategic Plan* with an intent to develop Health IT infrastructure and update EHRs' meaningful use criteria to include interoperability standards.[147]

Our findings also identified the need to determine the appropriate sample size of training data. As shown in Table 2, different studies have used different sample sizes (patients). How much data is sufficient to train an algorithm?—has been unanswered in the field of ML. Although ML performance generally improves with the additional information, plateaus exist wherein new information adds little to model performance.[148] In fact, some model's accuracy can be *hindered* with increasing information (data) usually because the additional variables tailor (overfit) the models for a too-specific set of information (context). Such model might perform poorly on new data, a problem long recognized as prediction bias or overfitting or minimal-optimal problem.[149] Nevertheless, this pursuit of the practical consideration results in another issue, known as the all-relevant problem,[150] which involves the identification of all attributes that are relevant for classification. The establishment of minimum data needs for adequate accuracy is required in healthcare. To determine these needs, we must understand factors that affect the amounts of data needed to achieve certain accuracy levels, an issue we refer to as *data efficiency*, including two components: (1) the rate at which accuracy increases with increasing data and (2) the maximum accuracy achievable by the method.[151]

The findings of our review, especially the limitations of ML, are in-line with the findings of other studies evaluating the impact of ML on different health care applications. A recent review by Battineni et al[152] analyzed the effect of ML on chronic disease diagnosis. It identified the dependence of ML on data and how different studies use different data set of varying data sizes to develop their ML model. Another review on the impact of AI on patient safety outcomes also identified the lack of AI standards.[153] Much work has been done in standardizing ML research and development. On February 11, 2019, the President (of the United States) issued an Executive Order (EO 13859) directing Federal agencies to develop a plan to ensure AI/ML standards.[154] A few months later, on June 17, 2019, China's Ministry of Science and Technology published a framework and action guidelines—*Principles for a New Generation of Artificial Intelligence: Develop Responsible Artificial Intelligence*.[155] In October 2019, the Office of the President of the Russian Federation released a national AI strategy. As of February 2020, there is also extensive information about Russian AI policy available that is published in OECD AI Policy Observatory.[156] The efforts taken so far in AI/ML standardization focus on developing a common (national) standard for all AI applications and ML algorithms. Since many of the issues around ML algorithms, particularly within healthcare, are context-specific, health care requires standards (governance) that are tailored toward its goal.[92,141]

## Relevance in clinical practices and recommendations

Besides the ML limitations and flaws discussed earlier, there are other limiting factors that can potentially inhibit the impact and growth of ML in geriatric care. Geriatric populations with multiple chronic complexities (MCC) are often excluded from randomized controlled trials.[157–160] Current approaches to geriatric guideline development usually emphasize single diseases, which may have minimal relevance to those with MCC.[161,162] Therefore it remains unclear which condition(s) contribute to an individual's health outcome, and consequently, which conditions should be the primary treatment target(s).[163] Unless the treatment target is specified, it gets challenging to implement ML models for diagnostic purposes that are trained for particular disease identification. This gives rise to a fundamental question of *whether it is appropriate or useful to identify a random ailment in a patient with MCC?* Consequently, *what is the applicability or importance of an ML model, trained for identifying single ailment, on a geriatric population with MCC?* In our review, all ML models for geriatric care were designed to diagnose or identify single ailment (Alzheimer's or depression or diabetic retinopathy). That would be interesting and needed future work to explore ML models considering MCC in their models for geriatrics patients.

Another consideration is that the available ML models, developed to help estimate prognosis, are based upon static data and algorithms. In contrast, a patient's health status is dynamic and changes over time. As a result, future research efforts to incorporate ML models into clinical workflow need to match the measure and underlying disease trajectory to the patient's individual situation.

ML studies often report their results in abstract measures like AUROC, F measure, etc. whereas, clinical trials or physicians typically evaluate and make decisions based on relative risk reduction (RRR) or absolute risk reduction (ARR).[164] Future research should develop an ML metric equivalent to ARR. ARR is often preferred to RRR because RRR is uninterpretable if the baseline risk is missing. The ARR is based on the risk of an outcome without treatment (or the baseline risk) minus the risk of the outcome with treatment. Studies implementing ML must consider the baseline risk for their outcome (diagnosis or recommendation) for patients with MCC (baseline risk for geriatric patient may be higher or lower than that of the general population).

ML models must also report the short-term, mid-term, and long-term effects of models' recommendations on patient health outcomes. When attempting to integrate ML-base prognosis into clinical decision-making, we recommend prioritizing decisions that are inclusive of life expectancy (short-term: within the next year; mid-term: within the next 5 years; long-term: beyond 5 years[165]). A patient with limited life expectancy would focus efforts on relevant short-term decisions, whereas patients with longer life expectancy might consider prognosis for mid-term or long-term care.

Although the science of ML-based prediction in medicine continues to evolve, some glitches concerning data exist. ML models or tools are usually developed and tested in specific settings, which potentially limit its measure's validity in other contexts. Often due to the quality of data or complexity of the algorithm, ML models might generate performance measures lower than the existing validated tools (expected). A study conducted by Marcantonio et al[166] recruited 201 geriatric patients and used *3D- Confusion Assessment Method* (3D-CAM) to evaluate psychological ailment with sensitivity of the 96% and specificity of 98%. Another study by Palmer et al recruited 30 patients and used 3-item questionnaire to identify AD with the maximum AUCROC of 0.97, and accuracy of 0.90. The study used *Mental State Examination (MMSE)* to identify AD with AUROC of 0.96.[167] Contrastingly, a study in our review used SVM and identified AD among healthy controls with an accuracy of 84.17%.[98] Another study used ECG data and artificial neural network to classify individuals with MCI that are likely to progress to AD with an accuracy of 85.98%.[101]

Due to the heterogeneity of geriatric patients and AI limitations such as (a) lack of AI standards, (b) lack of data governance, and (c) absence of an integrated global healthcare database, ML's integration into the clinical workflow will likely continue to challenge ML developers, clinicians, and policymakers.

### Limitation of the study

This study reviews publication that matches our inclusion criteria and operational definition of AI (ML). We also limited the scope of our review to geriatric patients with chronic conditions. Additionally, this review only includes studies published in the last 10 years in English.

## CONCLUSION

The results presented in this systematic review contributed to the understanding of the importance and use of AI in geriatric care. The review exhibits that ML algorithms were used to address many geriatrics diseases, which usually require just in time diagnosis and continuous care management by health care providers.

## SUPPLEMENTARY MATERIAL

Supplementary material is available at *Journal of the American Medical Informatics Association* online.

## AUTHOR CONTRIBUTIONS

O.A. conceived and designed the study, participated in data collection (literature review), analysis, and interpretation, drafted and revised the manuscript, and approved the final version for submission. A.C. participated in the literature review, analysis, and interpretation, prepared the graphical illustrations, drafted and revised the manuscript, and approved the final version for submission. E.R. participated in the literature review, analysis, and interpretation, drafted and revised the manuscript, and approved the final version for submission.

## FUNDING

This research received no specific grant from any funding agency in public, commercial, or not-for-profit sectors.

## CONFLICT OF INTEREST STATEMENT

None declared.

## REFERENCES

1. Vincent GK, Velkoff VA. *The next four decades: The older population in the United States: 2010 to 2050*. US Department of Commerce, Economics and Statistics Administration, US Census Bureau, 2010.
2. Miller KE, Zylstra RG, Standridge JB. The geriatric patient: a systematic approach to maintaining health. *Am Fam Physician* 2000; 61 (4): 1089–104.
3. Aungst RB. Healthy people 2020. *Perpect Audiol* 2011; 7 (1): 29–33.
4. Clerencia-Sierra M, Calderón-Larrañaga A, Martínez-Velilla N, *et al*. Multimorbidity patterns in hospitalized older patients: associations among chronic diseases and geriatric syndromes. *PLoS One* 2015; 10 (7): e0132909.
5. Marengoni A, Angleman S, Melis R, *et al*. Aging with multimorbidity: a systematic review of the literature. *Age Res Rev* 2011; 10 (4): 430–9.
6. Inouye SK, Studenski S, Tinetti ME, *et al*. Geriatric syndromes: clinical, research, and policy implications of a core geriatric concept: (see editorial comments by Dr. William Hazzard on pp 794–796). *J Am Geriatr Soc* 2007; 55 (5): 780–91.
7. Lee PG, Cigolle C, Blaum C. The co-occurrence of chronic diseases and geriatric syndromes: the Health and Retirement Study. *J Am Geriatr Soc* 2009; 57 (3): 511–6.
8. Gleberzon B. Chiropractic geriatrics: the challenges of assessing the older patient. *J Am Chiropractic Assoc* 2000; 43: 6–37.
9. Bowers LJ. Clinical assessment of geriatric patients: unique challenges. In: Mootz R, Bowers L, eds. Chiropractic Care of Special Populations. 1st ed. Geithersburg: Aspen Publisher; 2020: 168–170.
10. Killinger LZ, Morley JE, Kettner NW, *et al*. Integrated care of the older patient. *Topics Clin Chiropract* 2001; 8 (2): 46–55.
11. Hawk C, Schneider M, Dougherty P, *et al*. Best practices recommendations for chiropractic care for older adults: results of a consensus process. *J Manipulative Physiol Ther* 2010; 33 (6): 464–73.
12. Coulter ID, Hurwitz E, Aronow HU, Cassata D, Beck J. Chiropractic patients in a comprehensive home-based geriatric assessment, follow-up and health promotion program. Santa Monica, CA: RAND Corporation, RP-593; 1996.
13. Small GW, Rabins PV, Barry PP, *et al*. Diagnosis and treatment of Alzheimer disease and related disorders: consensus statement of the Ameri-can Association for Geriatric Psychiatry, the Alzheimer's Association, and the American Geriatrics Society. *JAMA* 1997; 278 (16): 1363–71.

14. Elsawy B, Higgins KE. The geriatric assessment. *Am Fam Physician* 2011; 83 (1): 48–56.

15. Tan T, Ong WS, Rajasekaran T, *et al.* Identification of comprehensive geriatric assessment based risk factors for malnutrition in elderly Asian cancer patients. *PLoS One* 2016; 11 (5): e0156008.

16. Enshaeifar S, Zoha A, Skillman S, *et al.* Machine learning methods for detecting urinary tract infection and analysing daily living activities in people with dementia. *PLoS One* 2019; 14 (1): e0209909.

17. Covinsky KE, Pierluissi E, Johnston CB. Hospitalization-associated disability: "She was probably able to ambulate, but I'm not sure". *JAMA* 2011; 306 (16): 1782–93.

18. Gill TM, Allore HG, Holford TR, *et al.* Hospitalization, restricted activity, and the development of disability among older persons. *JAMA* 2004; 292 (17): 2115–24.

19. Kansagara D, Englander H, Salanitro A, *et al.* Risk prediction models for hospital readmission: a systematic review. *JAMA* 2011; 306 (15): 1688–98.

20. Buurman BM, van Munster BC, Korevaar JC, *et al.* Prognostication in acutely admitted older patients by nurses and physicians. *J Gen Intern Med* 2008; 23 (11): 1883–9.

21. Walter LC, Brand RJ, Counsell SR, *et al.* Development and validation of a prognostic index for 1-year mortality in older adults after hospitalization. *JAMA* 2001; 285 (23): 2987–94.

22. Buurman BM, Hoogerduijn JG, de Haan RJ, *et al.* Geriatric conditions in acutely hospitalized older patients: prevalence and one-year survival and functional decline. *PLoS One* 2011; 6 (11): e26951.

23. Boyd CM, Landefeld CS, Counsell SR, *et al.* Recovery of activities of daily living in older adults after hospitalization for acute medical illness. *J Am Geriatr Soc* 2008; 56 (12): 2171–9.

24. Gill TM, Allore HG, Gahbauer EA, *et al.* Change in disability after hospitalization or restricted activity in older persons. *JAMA* 2010; 304 (17): 1919–28.

25. Covinsky KE, Palmer RM, Fortinsky RH, *et al.* Loss of independence in activities of daily living in older adults hospitalized with medical illnesses: increased vulnerability with age. *J Am Geriatr Soc* 2003; 51 (4): 451–8.

26. Boyd CM, Ricks M, Fried LP, *et al.* Functional decline and recovery of activities of daily living in hospitalized, disabled older women: the Women's Health and Aging Study I. *J Am Geriatr Soc* 2009; 57 (10): 1757–66.

27. Reichardt LA, Aarden JJ, van Seben R, *et al.*; on behalf of the Hospital-ADL study group. Unravelling the potential mechanisms behind hospitalization-associated disability in older patients; the Hospital-Associated Disability and impact on daily life (Hospital-ADL) cohort study protocol. *BMC Geriatr* 2016; 16 (1): 59.

28. Inouye SK, Charpentier PA. Precipitating factors for delirium in hospitalized elderly persons. Predictive model and interrelationship with baseline vulnerability. *JAMA* 1996; 275 (11): 852–7.

29. Sands LP, Yaffe K, Covinsky K, *et al.* Cognitive screening predicts magnitude of functional recovery from admission to 3 months after discharge in hospitalized elders. *J Gerontol A Biol Sci Med Sci* 2003; 58 (1): 37–45.

30. Reuben DB, Frank JC, Hirsch SH, *et al.* A randomized clinical trial of outpatient comprehensive geriatric assessment coupled with an intervention to increase adherence to recommendations. *J Am Geriatr Soc* 1999; 47 (3): 269–76.

31. Abbo ED, Zhang Q, Zelder M, *et al.* The increasing number of clinical items addressed during the time of adult primary care visits. *J Gen Intern Med* 2008; 23 (12): 2058–65.

32. Arndt BG, Beasley JW, Watkinson MD, *et al.* Tethered to the EHR: primary care physician workload assessment using EHR event log data and time-motion observations. *Ann Fam Med* 2017; 15 (5): 419–26.

33. Maly RC, Hirsch SH, Reuben DB. The performance of simple instruments in detecting geriatric conditions and selecting community-dwelling older people for geriatric assessment. *Age Ageing* 1997; 26 (3): 223–31.

34. Toseland RW, O'Donnell JC, Engelhardt JB, *et al.* Outpatient geriatric evaluation and management. Results of a randomized trial. *Med Care* 1996; 34 (6): 624–40.

35. Sarah M. Research Shows Shortage of More than 100,000 Doctors by 2030. Secondary Research Shows Shortage of More than 100,000 Doctors by 2030. 2017. https://www.aamc.org/news-insights/research-shows-shortage-more-100000-doctors-2030 Accessed June 2, 2020.

36. La Thangue NB, Kerr DJ. Predictive biomarkers: a paradigm shift towards personalized cancer medicine. *Nat Rev Clin Oncol* 2011; 8 (10): 587–96.

37. LeCun Y, Bengio Y, Hinton G. Deep learning. *Nature* 2015; 521 (7553): 436–44.

38. Mavaddat N, Michailidou K, Dennis J, *et al.* Polygenic risk scores for prediction of breast cancer and breast cancer subtypes. *Am J Hum Genet* 2019; 104 (1): 21–34.

39. Muttarak M, Peh WC, Euathrongchit J, *et al.* Spectrum of imaging findings in melioidosis. *BJR* 2009; 82 (978): 514–21.

40. O'Connor JP, Aboagye EO, Adams JE, *et al.* Imaging biomarker roadmap for cancer studies. *Nat Rev Clin Oncol* 2017; 14 (3): 169–86.

41. Wang M, Kaufman RJ. The impact of the endoplasmic reticulum protein-folding environment on cancer development. *Nat Rev Cancer* 2014; 14 (9): 581–97.

42. Wang X, Huang Y, Li L, *et al.* Assessment of performance of the Gail model for predicting breast cancer risk: a systematic review and meta-analysis with trial sequential analysis. *Breast Cancer Res* 2018; 20 (1): 18.

43. Caimmi M, Chiavenna A, Scano A, *et al.* Using robot fully assisted functional movements in upper-limb rehabilitation of chronic stroke patients: preliminary results. *Eur J Phys Rehabil Med* 2017; 53 (3): 390–9.

44. Cho KH, Song WK. Robot-assisted reach training for improving upper extremity function of chronic stroke. *Tohoku J Exp Med* 2015; 237 (2): 149–55.

45. Edwards DJ, Cortes M, Rykman-Peltz A, *et al.* Clinical improvement with intensive robot-assisted arm training in chronic stroke is unchanged by supplementary tDCS. *RNN* 2019; 37 (2): 167–80.

46. Khanna I, Roy A, Rodgers MM, *et al.* Effects of unilateral robotic limb loading on gait characteristics in subjects with chronic stroke. *J Neuroeng Rehabil* 2010; 7 (1): 23.

47. Nef T, Quinter G, Muller R, *et al.* Effects of arm training with the robotic device ARMin I in chronic stroke: three single cases. *Neurodegenerative Dis* 2009; 6 (5–6): 240–51.

48. Oude Nijeweme-d'Hollosy W, van Velsen L, Poel M, *et al.* Evaluation of three machine learning models for self-referral decision support on low back pain in primary care. *Int J Med Inform* 2018; 110: 31–41.

49. Ziherl J, Novak D, Olensek A, *et al.* Evaluation of upper extremity robot-assistances in subacute and chronic stroke subjects. *J Neuroeng Rehabil* 2010; 7 (1): 52.

50. Zollo L, Gallotta E, Guglielmelli E, *et al.* Robotic technologies and rehabilitation: new tools for upper-limb therapy and assessment in chronic stroke. *Eur J Phys Rehabil Med* 2011; 47 (2): 223–36.

51. Rodriguez-Gonzalez C, Herranz-Alonso A, Escudero-Vilaplana V, *et al.* Robotic dispensing improves patient safety, inventory management, and staff satisfaction in an outpatient hospital pharmacy. *J Eval Clin Pract* 2019; 25 (1): 28–35.

52. Zhao J, Wang G, Jiang Z, *et al.* Robotic gastrotomy with intracorporeal suture for patients with gastric gastrointestinal stromal tumors located at cardia and subcardiac region. *Surg Laparosc Endosc Percutan Tech* 2018; 28 (1): e1–7.

53. Denecke K. Automatic analysis of critical incident reports: requirements and use cases. *Stud Health Technol Inform* 2016; 223: 85–92.

54. Fong A, Howe J, Adams K, *et al.* Using active learning to identify health information technology related patient safety events. *Appl Clin Inform* 2017; 26 (1): 35–46.

55. Fong A, Ratwani R. An evaluation of patient safety event report categories using unsupervised topic modeling. *Methods Inf Med* 2015; 54 (4): 338–45.

56. Garcia-Jimenez A, Moreno-Conde A, Martinez-Garcia A, *et al.* Clinical decision support using a terminology server to improve patient safety. *Stud Health Technol Inform* 2015; 210: 150–4.

57. Kang H, Zhou S, Yao B, *et al.* A prototype of knowledge-based patient safety event reporting and learning system. *BMC Med Inform Decis Mak* 2018; 18 (S5): 110.

58. Kivekas E, Kinnunen UM, Paananen P, *et al.* Functionality of triggers for epilepsy patients assessed by text and data mining of medical and nursing records *Stud Health Technol Inform*. 2016; 225:128–32.

59. Li Y, Salmasian H, Harpaz R, *et al.* Determining the reasons for medication prescriptions in the EHR using knowledge and natural language processing. *AMIA Annu Symp Proc* 2011; 2011: 768–76.

60. McKnight SD. Semi-supervised classification of patient safety event reports. *J Patient Saf* 2012; 8 (2): 60–4.

61. Wang Y, Coiera E, Runciman W, *et al.* Using multiclass classification to automate the identification of patient safety incident reports by type and severity. *BMC Med Inform Decis Mak* 2017; 17 (1): 84.

62. Zhao J, Henriksson A, Asker L, *et al.* Predictive modeling of structured electronic health records for adverse drug event detection. *BMC Med Inform Decis Mak* 2015; 15 (S4): S1.

63. Dalal AK, Fuller T, Garabedian P, *et al.* Systems engineering and human factors support of a system of novel EHR-integrated tools to prevent harm in the hospital. *J Am Med Inform Assoc* 2019; 26 (6): 553–60.

64. Bahl M, Barzilay R, Yedidia AB, *et al.* High-risk breast lesions: a machine learning model to predict pathologic upgrade and reduce unnecessary surgical excision. *Radiology* 2018; 286 (3): 810–8.

65. Guan M, Cho S, Petro R, *et al.* Natural language processing and recurrent network models for identifying genomic mutation-associated cancer treatment change from patient progress notes. *JAMIA Open* 2019; 2 (1): 139–49.

66. Li Q, Zhao K, Bustamante CD, *et al.* Xrare: a machine learning method jointly modeling phenotypes and genetic evidence for rare disease diagnosis. *Genet Med* 2019; 21 (9): 2126–34.

67. Chen H, Engkvist O, Wang Y, *et al.* The rise of deep learning in drug discovery. *Drug Discovery Today* 2018; 23 (6): 1241–50.

68. Costabal FS, Matsuno K, Yao J, *et al.* Machine learning in drug development: characterizing the effect of 30 drugs on the QT interval using Gaussian process regression, sensitivity analysis, and uncertainty quantification. *Comput Methods Appl Mech Eng* 2019; 348: 313–33.

69. Ekins S, Puhl AC, Zorn KM, *et al.* Exploiting machine learning for end-to-end drug discovery and development. *Nat Mater* 2019; 18 (5): 435–41.

70. Jiang F, Jiang Y, Zhi H, *et al.* Artificial intelligence in healthcare: past, present and future. *Stroke Vasc Neurol* 2017; 2 (4): 230–43.

71. Banerjee I, Li K, Seneviratne M, *et al.* Weakly supervised natural language processing for assessing patient-centered outcome following prostate cancer treatment. *JAMIA Open* 2019; 2 (1): 150–9.

72. Ciervo J, Shen SC, Stallcup K, *et al.* A new risk and issue management system to improve productivity, quality, and compliance in clinical trials. *JAMIA Open* 2019; 2 (2): 216–21.

73. Ronquillo JG, Erik Winterholler J, Cwikla K, *et al.* Health IT, hacking, and cybersecurity: national trends in data breaches of protected health information. *JAMIA Open* 2018; 1 (1): 15–9.

74. Fogel AL, Kvedar JC. Artificial intelligence powers digital medicine. *NPJ Digit Med* 2018; 1: 5.

75. Galvin JE, Roe CM, Xiong C, *et al.* Validity and reliability of the AD8 informant interview in dementia. *Neurology* 2006; 67 (11): 1942–8.

76. Chin R, Ng A, Narasimhalu K, *et al.* Utility of the AD8 as a self-rating tool for cognitive impairment in an Asian population. *Am J Alzheimers Dis Other Demen* 2013; 28 (3): 284–8.

77. Shaik MA, Xu X, Chan QL, *et al.* The reliability and validity of the informant AD8 by comparison with a series of cognitive assessment tools in primary healthcare. *Int Psychogeriatr* 2016; 28 (3): 443–52.

78. Chan QL, Xu X, Shaik MA, *et al.* Clinical utility of the informant AD8 as a dementia case finding instrument in primary healthcare. *J Alzheimers Dis* 2015; 49 (1): 121–7.

79. Yang L, Yan J, Jin X, *et al.* Screening for dementia in older adults: comparison of Mini-Mental State Examination, Mini-Cog, Clock Drawing test and AD8. *PLoS One* 2016; 11 (12): e0168949.

80. Albert MV, Kording K, Herrmann M, *et al.* Fall classification by machine learning using mobile phones. *PLoS One* 2012; 7 (5): e36556.

81. Macrae C. Governing the safety of artificial intelligence in healthcare. *BMJ Qual Saf* 2019; 28 (6): 495–8.

82. Grossman LV, Choi SW, Collins S, *et al.* Implementation of acute care patient portals: recommendations on utility and use from six early adopters. *J Am Med Inform Assn* 2018; 25 (4): 370–9.

83. Chiu P-Y, Tang H, Wei C-Y, *et al.* NMD-12: a new machine-learning derived screening instrument to detect mild cognitive impairment and dementia. *PLoS One* 2019; 14 (3): e0213430.

84. Cabitza F, Rasoini R, Gensini GF. Unintended consequences of machine learning in medicine. *JAMA* 2017; 318 (6): 517–8.

85. Powers EM, Shiffman RN, Melnick ER, *et al.* Efficacy and unintended consequences of hard-stop alerts in electronic health record systems: a systematic review. *J Am Med Inform Assn* 2018; 25 (11): 1556–66.

86. McCarthy J, Hayes PJ. Some Philosophical Problems from the Standpoint of Artificial Intelligence. In: *Machine Intelligence*. Edinburgh: Edinburgh University Press; 1969: 463–502.

87. Hashimoto DA, Rosman G, Rus D, *et al.* Artificial intelligence in surgery: promises and perils. *Ann Surg* 2018; 268 (1): 70–6.

88. Bhardwaj R, Nambiar AR, Dutta D. A study of machine learning in healthcare. *P Int Comp Softw App* 2017: 236–41. doi:10.1109/Compsac.2017.164.

89. Kong H-J. Managing unstructured big data in healthcare system. *Healthc Inform Res* 2019; 25 (1): 1–2.

90. Lee RF, Lober WB, Sibley J, Kross EK, Engelberg RA, Curtis JR. Identifying goals-of-care conversations in the electronic health record using machine learning and natural language processing. In: *A22. FACILITATING PALLIATIVE AND END-OF-LIFE CARE*. American Thoracic Society; 2019; A1089.

91. CTA. Definitons and characteristics of artificial intelligence (ANSI/CTA-2089): Consumer Technology Association; 2020.

92. The Royal Society. *Machine Learning: The Power and Promise of Computers that Learn by Example: an Introduction*. Royal Society; 2017: 5–117.

93. Akl A, Taati B, Mihailidis A. Autonomous unobtrusive detection of mild cognitive impairment in older adults. *IEEE Trans Biomed Eng* 2015; 62 (5): 1383–94.

94. Al-Jarrah MA, Shatnawi H. Non-proliferative diabetic retinopathy symptoms detection and classification using neural network. *J Med Eng Technol* 2017; 41 (6): 498–505.

95. Andrews JA, Harrison RF, Brown LJE, *et al.* Using the NANA toolkit at home to predict older adults' future depression. *J Affect Disord* 2017; 213: 187–90.

96. Armananzas R, Bielza C, Chaudhuri KR, *et al.* Unveiling relevant non-motor Parkinson's disease severity symptoms using a machine learning approach. *Artif Intell Med* 2013; 58 (3): 195–202.

97. Ashraf A, Taati B. Automated video analysis of handwashing behavior as a potential marker of cognitive health in older adults. *IEEE J Biomed Health Inform* 2016; 20 (2): 682–90.

98. Beheshti I, Maikusa N, Daneshmand M, *et al.*; for the Japanese-Alzheimer's Disease Neuroimaging Initiative. Classification of Alzheimer's disease and prediction of mild cognitive impairment conversion using histogram-based analysis of patient-specific anatomical brain connectivity networks. *J Alzheimers Dis* 2017; 60 (1): 295–304.

99. Bhagwat N, Viviano JD, Voineskos AN, *et al.*; Alzheimer's Disease Neuroimaging Initiative. Modeling and prediction of clinical symptom trajectories in Alzheimer's disease using longitudinal data. *PLoS Comput Biol* 2018; 14 (9): e1006376.

100. Bogunovic H, Montuoro A, Baratsits M, *et al.* Machine learning of the progression of intermediate age-related macular degeneration based on OCT imaging. *Invest Ophthalmol Vis Sci* 2017; 58 (6): Bio141.

101. Buscema M, Grossi E, Capriotti M, *et al.* The I.F.A.S.T. model allows the prediction of conversion to Alzheimer disease in patients with mild cognitive impairment with high degree of accuracy. *Curr* 2010; 7 (2): 173–87.

102. Cabral C, Morgado PM, Campos Costa D, *et al.* Predicting conversion from MCI to AD with FDG-PET brain images at different prodromal stages. *Comput Biol Med* 2015; 58: 101–9.

103. Cheng B, Liu M, Suk HI, *et al.*; Alzheimer's Disease Neuroimaging Initiative. Multimodal manifold-regularized transfer learning for MCI conversion prediction. *Brain Imaging Behav* 2015; 9 (4): 913–26.

104. Choi J, Choi J, Choi WJ. Predicting depression among community residing older adults: a use of machine learning approach. *Stud Health Technol Inform* 2018; 250: 265.

105. Christopher M, Belghith A, Weinreb RN, *et al*. Retinal nerve fiber layer features identified by unsupervised machine learning on optical coherence tomography scans predict glaucoma progression. *Invest Ophthalmol Vis Sci* 2018; 59 (7): 2748–56.

106. Dias A, Gorzelniak L, Schultz K, *et al*. Classification of exacerbation episodes in chronic obstructive pulmonary disease patients. *Methods Inf Med* 2014; 53 (2): 108–14.

107. Du Y, Almajalid R, Shan J, *et al*. A novel method to predict knee osteoarthritis progression on MRI using machine learning methods. *IEEE Transon Nanobioscience* 2018; 17 (3): 228–36.

108. Goldbaum MH, Lee I, Jang G, *et al*. Progression of patterns (POP): a machine classifier algorithm to identify glaucoma progression in visual fields. *Invest Ophthalmol Vis Sci* 2012; 53 (10): 6557–67.

109. Hatton CM, Paton LW, McMillan D, *et al*. Predicting persistent depressive symptoms in older adults: a machine learning approach to personalised mental healthcare. *J Affect Disord* 2019; 246: 857–60.

110. Hubenthal M, Hemmrich-Stanisak G, Degenhardt F, *et al*. Sparse modeling reveals miRNA signatures for diagnostics of inflammatory bowel disease. *PLoS One* 2015; 10 (10): e0140155.

111. James SL, Romero M, Ramírez-Villalobos D, *et al*. Validating estimates of prevalence of non-communicable diseases based on household surveys: the symptomatic diagnosis study. *BMC Med* 2015; 13 (1): 15.

112. Konig A, Linz N, Zeghari R, *et al*. Detecting apathy in older adults with cognitive disorders using automatic speech analysis. *J Alzheimers Dis* 2019; 69 (4): 1183–93.

113. Kuo KM, Talley PC, Huang CH, *et al*. Predicting hospital-acquired pneumonia among schizophrenic patients: a machine learning approach. *BMC Med Inform Decis Mak* 2019; 19 (1): 42.

114. Lara J, López-Labrador F, González-Candelas F, *et al*. Computational models of liver fibrosis progression for hepatitis C virus chronic infection. *BMC Bioinformatics* 2014; 15 (S8): S5.

115. Lee G, Nho K, Kang B, *et al*.; for Alzheimer's Disease Neuroimaging Initiative. Predicting Alzheimer's disease progression using multi-modal deep learning approach. *Sci Rep* 2019; 9 (1): 1952.

116. Lins A, Muniz MTC, Garcia ANM, *et al*. Using artificial neural networks to select the parameters for the prognostic of mild cognitive impairment and dementia in elderly individuals. *Comput Methods Programs Biomed* 2017; 152: 93–104.

117. Morabito FC, Campolo M, Mammone N, *et al*. Deep learning representation from electroencephalography of early-stage Creutzfeldt-Jakob disease and features for differentiation from rapidly progressive dementia. *Int J Neur Syst* 2017; 27 (2): 1650039.

118. Naushad SM, Hussain T, Indumathi B, *et al*. Machine learning algorithm-based risk prediction model of coronary artery disease. *Mol Biol Rep* 2018; 45 (5): 901–10.

119. Ortiz A, Munilla J, Gorriz JM, *et al*. Ensembles of deep learning architectures for the early diagnosis of the Alzheimer's disease. *Int J Neur Syst* 2016; 26 (7): 1650025.

120. Park H, Yang JJ, Seo J, *et al*. Dimensionality reduced cortical features and their use in predicting longitudinal changes in Alzheimer's disease. *Neurosci Lett* 2013; 550: 17–22.

121. Pedoia V, Haefeli J, Morioka K, *et al*. MRI and biomechanics multidimensional data analysis reveals R2 -R1rho as an early predictor of cartilage lesion progression in knee osteoarthritis. *J Magn Reson Imaging* 2018; 47 (1): 78–90.

122. Schmidt-Erfurth U, Waldstein SM, Klimscha S, *et al*. Prediction of individual disease conversion in early AMD using artificial intelligence. *Invest Ophthalmol Vis Sci* 2018; 59 (8): 3199–208.

123. Thung KH, Wee CY, Yap PT, *et al*. Identification of progressive mild cognitive impairment patients using incomplete longitudinal MRI scans. *Brain Struct Funct* 2016; 221 (8): 3979–95.

124. Treder M, Lauermann JL, Eter N. Deep learning-based detection and classification of geographic atrophy using a deep convolutional neural network classifier. *Graefes Arch Clin Exp Ophthalmol* 2018; 256 (11): 2053–60.

125. Weakley A, Williams JA, Schmitter-Edgecombe M, *et al*. Neuropsychological test selection for cognitive impairment classification: a machine learning approach. *J Clin Exp Neuropsychol* 2015; 37 (9): 899–916.

126. Yang ST, Lee JD, Chang TC, *et al*. Discrimination between Alzheimer's disease and mild cognitive impairment using SOM and PSO-SVM. *Comput Math Methods Med* 2013; 2013: 1.

127. Zhan Y, Chen K, Wu X, *et al*.; for the Alzheimer's Disease Neuroimaging Initiative1. Identification of conversion from normal elderly cognition to Alzheimer's disease using multimodal support vector machine. *J Alzheimers Dis* 2015; 47 (4): 1057–67.

128. Luchi RJ, Gammack JK, Victor J, Narcisse I. Standards of care in geriatric practice. *Annu Rev Med* 2003; 54 (1): 185–96.

129. Thabtah F. Machine learning in autistic spectrum disorder behavioral research: a review and ways forward. *Inform Health Soc Care* 2019; 44 (3): 278–97.

130. Bhhatarai B, Walters WP, Hop C, *et al*. Opportunities and challenges using artificial intelligence in ADME/Tox. *Nat Mater* 2019; 18 (5): 418–22.

131. Berner ES, Ozaydin B. Benefits and risks of machine learning decision support systems. *JAMA* 2017; 318 (23): 2353–4.

132. ISO. ISO/IEC CD 23053 framework for artificial intelligence (AI) systems using machine learning (ML). Framework for artificial intelligence (AI) systems using machine learning (ML); 2020. https://www.iso.org/standard/74438.html?browse=tc

133. ISO. ISO/WD TTR 22100-5. Safety of machinery—Relationship with ISO 12100—Part 5: Implications of embedded Artificial Intelligence-machine learning; 2020. https://www.iso.org/standard/80778.html

134. ISO. ISO/IEC JTC 1/SC 42. Artificial intelligence; 2017. https://www.iso.org/committee/6794475.html

135. Huang J, Ling CX. Using AUC and accuracy in evaluating learning algorithms. *IEEE Trans Knowledge Data Eng* 2005; 17 (3): 299–310.

136. Stafford IS, Kellermann M, Mossotto E, *et al*. A systematic review of the applications of artificial intelligence and machine learning in autoimmune diseases. *NPJ Digit Med* 2020; 3 (1): 30.

137. Kurczab R, Smusz S, Bojarski AJ. The influence of negative training set size on machine learning-based virtual screening. *J Cheminform* 2014; 6 (1): 32.

138. Davis J, Goadrich M. The relationship between Precision-Recall and ROC curves. In: proceedings of the 23rd international conference on Machine learning (ICML '06). New York, NY: Association for Computing Machinery; 2006: 233–40.

139. Parikh RB, Obermeyer Z, Navathe AS. Regulation of predictive analytics in medicine. *Science* 2019; 363 (6429): 810–2.

140. Gianfrancesco MA, Tamang S, Yazdany J, *et al*. Potential biases in machine learning algorithms using electronic health record data. *JAMA Intern Med* 2018; 178 (11): 1544–7.

141. Michael M, Sonoo I, Mahnoorr A, *et al*. *Artificial Intelligence in Health Care: The Hope, the Hype, the Promise, the Peril*. Washington, DC: National Academy of Medicine, 2019.

142. Davatzikos C. Machine learning in neuroimaging: progress and challenges. *Neuroimage* 2019; 197: 652–6.

143. Maddox TM, Rumsfeld JS, Payne P. Questions for artificial intelligence in health care. *JAMA* 2019; 321 (1): 31–2.

144. Popejoy AB, Fullerton SM. Genomics is failing on diversity. *Nature News* 2016; 538 (7624): 161–4.

145. Bustamante CD, De La Vega FM, Burchard EG. Genomics for the world. *Nature* 2011; 475 (7355): 163–5.

146. Hofer IS, Burns M, Kendale S, Wanderer JP. Realistically integrating machine learning into clinical practice. A road map of opportunities, challenges, and a potential future. *Anesth Analg* 2020; 130 (5): 1115–8.

147. Rucker DW. 2020-2025 Federal health IT strategic plan. In: Schneider J, ed. Technology TOotNCfHI, ed. Draft for public comment ed. Texas Medical Association; 2020: 18.

148. Stockwell DRB, Peterson AT. Effects of sample size on accuracy of species distribution models. *Ecol Model* 2002; 148 (1): 1–13.

149. Choudhury A, Eksioglu B. Using predictive analytics for cancer identification. In: Romeijn HE, Schaefer A, Thomas R, eds. *Proceedings of the IISE Annual Conference and Expo*. Orlando: IISE; 2019: 1–7.

150. Kursa MB, Rudnicki WR. Feature selection with the Boruta package. *J Stat Softw* 2010; 36 (11): 1–13.

151. Stockwell DR. Generic predictive systems: an empirical evaluation using the Learning Base System (LBS). *Expert Syst Appl* 1997; 12 (3): 301–10.

152. Battineni G, Sagaro GG, Chinatalapudi N. Applications of machine learning predictive models in the chronic disease diagnosis. *J Pers Med* 2020; 10 (2): E21.

153. Choudhury A, Asan O. Role of artificial intelligence in patient safety outcomes: systematic literature review. *JMIR Med Informatics* 2020; 8 (7): e18599.

154. Nist US. Leadership in AI: a plan for federal engagement in developing technical standards and related tools. U.S. Department of Commerce; 2019: 1–45. https://www.nist.gov/system/files/documents/2019/08/10/ai_standards_fedengagement_plan_9aug2019.pdf

155. Laskai L, Websteer G. Governance Principles for a New Generation of Artificial Intelligence: Develop Responsible Artificial Intelligence: National New Generation Artificial Intelligence Governance Expert Committee; 2020. http://chinainnovationfunding.eu/dt_testimonials/publication-of-the-new-generation-ai-governance-principles-developing-responsible-ai/

156. OECD.AI. Policy Initiatives for Russian Federation. Secondary Policy Initiatives for Russian Federation; 2020. https://oecd.ai/dashboards/policy-initiatives?conceptUris=http.%2F%2Fkim.oecd.org%2FTaxonomy%2FGeographica-lAreas%23RussianFederation Accessed June 2, 2020.

157. Boyd CM, Kent DM. Evidence-based medicine and the hard problem of multimorbidity. *J Gen Intern Med* 2014; 29 (4): 552–3.

158. Uhlig K, Leff B, Kent D, *et al*. A framework for crafting clinical practice guidelines that are relevant to the care and management of people with multimorbidity. *J Gen Intern Med* 2014; 29 (4): 670–9. Online First.

159. Weiss CO, Varadhan R, Puhan MA, *et al*. Multimorbidity and evidence generation. *J Gen Intern Med* 2014; 29 (4): 653–60.

160. Navar AM, Pencina MJ, Peterson ED. ED. Assessing cardiovascular risk to guide hypertension diagnosis and treatment. *JAMA Cardiol* 2016; 1 (8): 864–71.

161. Trikalinos TA, Segal JB, Boyd CM. Addressing multimorbidity in evidence integration and synthesis. *J Gen Intern Med* 2014; 29 (4): 661–9.

162. Fabbri LM, Boyd C, Boschetto P, *et al*. How to integrate multiple comorbidities in guideline development: article 10 in integrating and coordinating efforts in COPD guideline development. An official ATS/ERS workshop report. *Proc Am Thorac Soc* 2012; 9 (5): 274–81.

163. Tinetti ME, McAvay GJ, Chang SS, *et al*. Contribution of multiple chronic conditions to universal health outcomes. *J Am Geriatr Soc* 2011; 59 (9): 1686–91.

164. Society AG. Guiding principles for the care of older adults with multimorbidity: an approach for clinicians: American Geriatrics Society Expert Panel on the Care of Older Adults with Multimorbidity. *J Am Geriatr Soc* 2012; 60 (10): E1–25.

165. Yourman LC, Lee SJ, Schonberg MA, *et al*. Prognostic indices for older adults: a systematic review. *JAMA* 2012; 307 (2): 182–92.

166. Marcantonio ER, Ngo LH, O'Connor M, *et al*. 3D-CAM: derivation and validation of a 3-minute diagnostic interview for CAM-defined delirium: a cross-sectional diagnostic test study. *Ann Intern Med* 2014; 161 (8): 554–61.

167. Palmer BW, Dunn LB, Appelbaum PS, *et al*. Assessment of capacity to consent to research among older persons with schizophrenia, Alzheimer disease, or diabetes mellitus: comparison of a 3-item questionnaire with a comprehensive standardized capacity instrument. *Arch Gen Psychiatry* 2005; 62 (7): 726–33.